\documentclass[12pt]{article}
\usepackage{amsfonts,psfig,epsfig,latexsym,amscd,multicol,theorem,}
\newcommand{\R}{{\mathbb{R}}} 
\newcommand{\Z}{{\mathbb{Z}}}

\newcommand{\beq}{\begin{equation}} 
\newcommand{\eeq}{\end{equation}} 
\newcommand{\bea}{\begin{eqnarray}} 
\newcommand{\eea}{\end{eqnarray}} 
\newcommand{\ra}{\rightarrow}

\newcommand{\cd}{\partial} 
 
\newcommand{\wh}{\widehat} 
 
\newcommand{\eps}{{\varepsilon}}

\newcommand{\hess}{{\sf Hess}} 
\newcommand{\spec}{{\rm spec\, }}

\newcommand{\dv}{{\bf d}} 
\newcommand{\nv}{{\bf n}} 
\newcommand{\rv}{{\bf r}} 
\newcommand{\veps}{{\varepsilon}}

\newcommand{\mod}{{\rm mod}\, } 
 
\theoremstyle{plain} 
\newtheorem{thm}{Theorem} 
\newtheorem{lemma}[thm]{Lemma} 
\newtheorem{prop}[thm]{Proposition} 
\newtheorem{cor}[thm]{Corollary} 
 {\theorembodyfont{\rmfamily} 
\newtheorem{defn}[thm]{Definition}

} 
\newcommand{\news}{\setcounter{equation}{0}} 
\newcommand{\comment}[1]{}

\begin{document} 

\title{Kinks in dipole chains} 
\author{
J.M. Speight\thanks{E-mail: {\tt speight@maths.leeds.ac.uk}} \\
School of Mathematics, University of Leeds\\
Leeds LS2 9JT, England\\ \\
Y. Zolotaryuk\thanks{E-mail: {\tt yzolo@bitp.kiev.ua}} \\
Bogolyubov Institute for Theoretical Physics \\
National Academy of Sciences of Ukraine\\
Kiev 03143, Ukraine
} 

\date{} 

\maketitle 

\begin{abstract}
It is shown that the topological discrete sine-Gordon system 
introduced by Speight
and Ward models the dynamics of an infinite uniform chain of electric
dipoles constrained to rotate in a plane containing the chain. Such
a chain admits a novel type of static kink solution 
which may occupy any position
relative to the spatial lattice and 
experiences no Peierls-Nabarro
barrier. Consequently the dynamics of a single kink is highly
continuum like, despite the strongly discrete nature of the model. 
Static multikinks and kink-antikink 
pairs are constructed, and it is shown that
all such static solutions are unstable. Exact propagating kinks are
sought numerically
using the pseudo-spectral method, but it is found that none exist,
except, perhaps, at very low speed. 

\begin{flushright}PACS Classification: 05.45.-a, 05.45.Yv, 05.50.+q
\end{flushright}
\end{abstract}

\section{Introduction}\label{sec:intro} 
\news

Nonlinear Klein-Gordon kinks are found in many branches of theoretical
physics. For applications in condensed matter and biophysics, the 
nonlinear Klein-Gordon equation is usually a continuum approximation of
a more fundamental spatially discrete system: kinks really live on a lattice
reflecting some crystalline or biomolecular structure. It has long been
recognized that spatial discreteness introduces important new effects  into
the dynamics of kinks
\cite{peykru}. Most notable is the Peierls-Nabarro (PN) barrier. Due
to the loss of continuous translation symmetry, static kinks may
generically be centred only on a lattice site or exactly half way between
lattice sites. The PN barrier is the energy difference between these two
solutions. A kink moving with too little kinetic energy cannot surmount the
PN barrier and becomes trapped. A kink moving with high kinetic energy
loses energy by emitting radiation and may also, eventually, become trapped.
So the dynamics of discrete kinks is much more complicated than their 
continuum counterparts.

This rich behaviour has led mathematical physicists to study discrete
Klein-Gordon systems extensively for their own sake, independent
of any physical application. In this regard, there has been a recent
resurgence of interest in so-called {\em exceptional}\, discretizations of
nonlinear Klein-Gordon models: those which, despite the loss of
translation symmetry, maintain a continuous family of static kink solutions,
centred anywhere relative to the lattice (loosely, a continuous translation 
orbit of static kinks). An early example of this was the topological
discrete sine-Gordon (TDSG) system \cite{spewar}, which eliminated the
PN barrier by preserving the Bogomol'nyi bound on kink energy. The idea was
subsequently applied to the $\phi^4$ model \cite{spe2} and then generalized 
to all Klein-Gordon models \cite{spe3}. A rather different approach is to
choose the continuous family of static kinks in advance, 
then reverse-engineer a
discrete Klein-Gordon system which supports them \cite{flaklazol}, the 
so-called inverse method. Both this method and the topological discretization
approach discretize the Lagrangian of the system, so they preserve an
energy conservation law. Kevrekidis has given a method for constructing
exceptional discretizations of the PDE itself, which destroys the variational
set-up, but preserves a {\em momentum}\, conservation law \cite{kev}.  
With collaborators, he
has recently shown that, if one discretizes the Laplacian in standard 
fashion,
it is impossible to preserve both energy and momentum conservation
\cite{dmikevyos}. Note that
in the absence of a conserved energy it is not meaningful to speak of a PN
potential. Most recently, Barashenkov, Oxtoby and Pelinovski \cite{baroxtpel}
have 
systematically
constructed all polynomial
exceptional discretizations of the $\phi^4$ PDE with
standard discrete Laplacian, finding families which generalize the 
discrete $\phi^4$ systems of
\cite{spe2} and  \cite{kev}. The new exceptional discretizations they find
preserve neither energy nor momentum conservation. Moving outside the
polynomial class, Dmitriev, Kevrekidis and Yoshikawa have constructed
yet more exceptional discrete Klein-Gordon models, concentrating on
the $\phi^4$ case, some of which conserve energy, some momentum, but none
both \cite{dmikevyos2}.

All these systems are theoretically interesting, but from a physical 
viewpoint they look rather contrived, and it is hard to imagine a concrete
physical system that they model. The purpose of this paper is to point out
that one of them does, in fact, model a simple physical system:
 a uniform chain of electric dipoles. The kinks interpolate between the
two vacuum states where the dipoles are aligned uniformly forward and 
uniformly backward along the chain. This system has three parameters:
the dipole strength $d$, the inter-dipole separation $\delta$ and the
dipole moment of inertia $I$. For all values of these parameters, this 
system can be reduced to the TDSG system at a single fixed value of the 
lattice spacing ($h=\sqrt{12}$, it turns out). This allows a concrete
physical reinterpretation of much of the previous work on the TDSG system
\cite{spewar,kot,hasspe,spebre}. 
It also motivates us to investigate the TDSG system
more deeply, yielding some new results. We will construct static multikink
and kink-antikink pair solutions, and prove that all static solutions apart 
from the kink, antikink and vacua are unstable. We will also numerically
seek exact propagating kinks using the pseudo-spectral method, but
will find only {\it nanopterons}: travelling kinks with spatially
oscillatory tails. The tail amplitude becomes extremely small for
low propagation speed, so that the possibility of exact propagating
low speed kinks cannot yet be discounted.

In section \ref{sec:dip}, we will introduce the dipole chain and demonstrate
that it is modelled by the TDSG system. In section \ref{sec:stat} we will
construct static solutions with arbitrary winding number and prove that
all ``non-Bogomol'nyi'' static solutions (that is, all solutions except 
kinks, antikinks and vacua) are unstable. Section \ref{sec:prop}
presents our results on exact propagating solutions, 
while section \ref{sec:conc} contains some concluding remarks.

\section{Dipole chains}\label{sec:dip}
\news

If two electric dipoles of moments $\dv_1$, $\dv_2$ are held
at positions $\rv_1$, $\rv_2\in\R^3$,  the potential energy due to their
electrostatic interaction is
\beq\label{dipdip}
\delta E=\frac{1}{4\pi\veps_0}\frac{\dv_1\cdot\dv_2-3(\nv\cdot\dv_1)(\nv\cdot
\dv_2)}{|\rv_1-\rv_2|^3},
\eeq
where $\nv=(\rv_1-\rv_2)/|\rv_1-\rv_2|$. Consider an infinite chain of
equally spaced dipoles, all of equal strength, each free to rotate in some
fixed plane containing the axis of the chain. Without loss of generality
we can assume the chain is directed along the $x$-axis and the dipoles
rotate in the $(x,y)$-plane. Then the $n$th dipole is located at
$\rv_n=(\delta n,0,0)$ 
and has moment $\dv_n=d(\cos\phi_n,\sin\phi_n,0)$, where
$\delta$ is the inter-dipole spacing and $\phi_n$ 
is the angle between $\dv_n$ and the $x$-axis. Hence, the 
potential energy 
of a neighbouring pair $\dv_n$, $\dv_{n+1}$ is
\bea
\delta E_n&=&\frac{d^2}{4\pi\veps_0\delta^3}\left[\cos(\phi_{n+1}-\phi_n)-
3\cos\phi_n\cos\phi_{n+1}\right]\nonumber\\
&=&\frac{d^2}{4\pi\veps_0\delta^3}\left[\sin^2\frac{1}{2}(\phi_{n+1}-\phi_n)
+3\sin^2\frac{1}{2}(\phi_{n+1}+\phi_n)-2\right].
\eea
It is convenient to add $d^2/2\pi\veps_0\delta^3$ to $\delta E_n$ so that the
minimum energy of a neighbouring pair is normalized to $0$, occurring
when $\dv_{n}=\dv_{n+1}=(\pm d,0,0)$. The total potential energy of the
chain, neglecting longer range interactions, is
\beq
E_P=\frac{d^2}{4\pi\veps_0\delta^3}\sum_{n\in\Z}\left\{
\sin^2\frac{1}{2}(\phi_{n+1}-\phi_n)
+3\sin^2\frac{1}{2}(\phi_{n+1}+\phi_n)\right\},
\eeq
while its total kinetic energy is
\beq
E_K=\sum_{n\in\Z}\frac{1}{2}\, I\, \left(\frac{d\phi_n}{dt}\right)^2,
\eeq
where $I$ is the moment of inertia of a dipole. The Lagrangian
governing its time evolution is then $L=E_K-E_P$.

If we define the natural units of energy and time to be 
$E_0=\sqrt{3}d^2/2\pi\veps_0\delta^3$ and $T_0=(I/\sqrt{3} E_0)^\frac{1}{2}$
respectively,
this Lagrangian coincides with that of the topological discrete sine-Gordon
system \cite{spewar}
with lattice spacing $\sqrt{12}$. More precisely, let us
define rescaled energy and time variables $\wh{E}_K=E_K/E_0$, 
$\wh{E}_P=E_P/E_0$ and $\wh{t}=t/T_0$. Then
\beq
\wh{L}=\wh{E}_K-\wh{E}_P=\frac{h}{4}\sum_n\left\{
\dot{\phi}_n^2-\left(\frac{2}{h}\sin\frac{1}{2}(\phi_{n+1}-\phi_n)\right)^2
-\sin^2\frac{1}{2}(\phi_{n+1}+\phi_n)\right\}
\eeq
where $h=\sqrt{12}$ and $\dot{f}:=df/d\wh{t}$. This is precisely the
TDSG Lagrangian. It is convenient to take $\delta$ as the unit of
length.
Note that $\delta,d,I$ affect the energy and length scales of the system, but
not its dynamical properties. These depend only on $h$, which is
fixed at $\sqrt{12}$ for all dipole chains. Note also that the physical
lattice spacing of the chain is $\delta$, not $h$.
We shall henceforth use the rescaled variables exclusively, and
drop the\,\, $\wh{}$\,\,  superscripts.

The equation of motion of the TDSG system is
\beq\label{eom}
\ddot{\phi}_n=\frac{4-h^2}{4h^2}\cos\phi_n(\sin\phi_{n+1}+\sin\phi_{n-1})
-\frac{4+h^2}{4h^2}\sin\phi_n(\cos\phi_{n+1}+\cos\phi_{n-1}).
\eeq
If $0<h<2$, it supports static kink solutions
\beq\label{kink}
\phi_n=2\tan^{-1}e^{a(n-b)},\qquad
a=\log\left(\frac{2+h}{2-h}\right),
\eeq
centred anywhere on the lattice (the kink position $b$ may take any real
value). These kinks all have energy $1$, the minimum possible for a 
configuration with kink boundary behaviour ($\lim_{n\ra\pm\infty}\phi_n=
 k_\pm\pi$ with $|k_+-k_-|=1$). Thus they experience
no PN barrier -- this is an exceptional discretization of the continuum
sine-Gordon model, which we recover in the limit $h\ra 0$. 
Note that adding any integer multiple
of $2\pi$ to a single $\phi_n$ preserves solutions of (\ref{eom}), so one
should think of $\phi_n$ as an angular coordinate of period $2\pi$. 
The
constant sequences $\phi_n=0$ and $\phi_n=\pi$ are two distinct vacuum
solutions, between which the kink (\ref{kink}) interpolates. The dynamics of
a single kink was studied in detail in \cite{spewar}. It was found that
kinks may propagate with arbitrarily low speed and never get pinned. Fast
kinks on coarse lattices ($h$ close to $2$) do excite significant
radiation in their wake which causes them to decelerate appreciably, but 
this effect is far weaker than in the conventional discrete system with
the same $h$. We shall return to the question of whether the system 
admits {\em exact} propagating kinks with constant speed in section
\ref{sec:prop}, but in any case kinks are very highly mobile in this
system. Kink-antikink collisions were studied numerically by Kotecha in
\cite{kot}, in the case where the kink and antikink are fired at one another
with equal speed. Such collisions
 were found to be rather similar to continuum $\phi^4$
kink-antikink collisions. There is a critical speed
$v_c$, depending on $h$, above which
the kink and antikink pass through one another. Kotecha found that
$v_c$, measured in lattice cells traversed per unit time,
depends approximately linearly on $h$. Below $v_c$ there is a
complicated (possibly fractal) structure of velocity windows in which the
kink and antikink ``bounce'' off one another $B\geq 1$ times, where a
bounce consists of the kink passing through the antikink, then turning
around and passing back through in the opposite direction. There seem to be
windows with $B=\infty$, meaning that the kink-antikink pair settles into
a breather-like oscillatory bound state. In fact, the TDSG 
system supports exact
breather solutions \cite{hasspe}, at least for $h$ close to $2$, 
though these 
are all unstable \cite{spebre}. 

\begin{figure}[htb]
\begin{center}\includegraphics[scale=0.8]{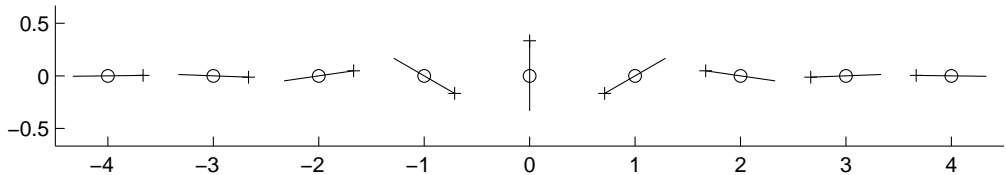}\end{center}
\caption{A site centred dipole kink ($b=0$).}\label{fig1}
\end{figure}

In the case of the dipole chain, $h=\sqrt{12}>2$, so the results of
\cite{spewar,kot,hasspe,spebre} 
do not directly apply. To make use of them we must exploit
a symmetry of (\ref{eom}): $\phi_n(t)$ is a solution of the
spacing $h$
system  if and only if 
\beq\label{sym}
\psi_n(t)=(-1)^n\phi_n(2t/h)
\eeq
is a solution of the spacing $h_*=4/h$ system. Note that if $h>2$ then
$0<h_*<2$. We can thus identify the dipole chain with the TDSG system with
lattice spacing $h_*=4/\sqrt{12}=2/\sqrt{3}<2$, so the dipole chain
supports a continuous translation orbit of ``alternating'' kinks
\beq
\label{akink}
\phi_n=(-1)^n2\tan^{-1}e^{a(n-b)},\qquad
a=\log\left(\frac{\sqrt{3}+1}{\sqrt{3}-1}\right),
\eeq
parametrized by $b\in\R$, all of which have energy $1$. Note that
$\phi_n$ still has kink boundary behaviour because $\phi_n$ is angular,
so $\pi$ and $-\pi$ are identified. The kink experiences no PN barrier.
These kinks are very highly discrete, in that their structure is spread
over very few lattice sites. For example, $86.6$\% of the energy of the
$b=0$ kink resides in the $-1\leq n\leq 1$ section of the chain. This
kink is depicted in figure \ref{fig1}.

Since $h_*\approx 1.15$ is close to $1$, the numerical results of
\cite{spewar} show that its propagation at modest speed
is strongly continuum like: radiative deceleration is weak and the kink
speed varies very little over a single lattice cell transit. Like their
$h<2$ counterparts, therefore, kinks in dipole chains are highly mobile.
Radiative deceleration is essentially negligible if the frequency with
which the kink centre hits lattice sites is below the bottom edge of the
phonon band of the lattice \cite{spewar}. In the case of dipole kinks, this 
happens when
\beq
2\pi v<\frac{2}{h}=\frac{1}{\sqrt{3}}\quad\Leftrightarrow\quad
v<\frac{1}{2\pi\sqrt{3}}\approx 0.092\, \mbox{sites/unit time.}
\eeq
In ``laboratory'' units, this becomes 
\beq
v<\frac{d}{\sqrt{8\pi^3\veps_0I\delta}}.
\eeq
Turning to kink-antikink collisions, we can estimate the critical incidence
speed, above which they pass through one another, for the spacing
$h_*=2/\sqrt{3}$ lattice, using Kotecha's
linear fit: $v_c\approx 0.120$ sites/unit time \cite{kot}. From (\ref{sym})
it then follows that the critical speed for dipole kink-antikink collisions
is
\beq
v_c\approx 0.208\,\mbox{sites/unit time}\approx 0.144
 \frac{d}{\sqrt{\veps_0I\delta}}.
\eeq
The numerical results of \cite{hasspe} show that $h_*$ is too far from
$2$ for the dipole chain to support any breather solution constructed
by continuation from the anti-continuum limit. 

\section{Static solutions}
\news
\label{sec:stat}

An important consequence of the PN potential in conventional discrete
systems is that they can, 
in contrast to their continuum counterparts, support finite
energy static multikink solutions. The point is
that if two kinks are well separated, their mutual repulsion is too weak
to overcome the pinning force due to the PN barrier. One can thus
construct static solutions with arbitrarily many kinks and antikinks in
stable equilibrium. Since the TDSG system has no PN barrier one might 
expect it to have no static multikinks either. However, this turns out to be
false: one can construct static solutions with $\lim_{n\ra\pm\infty}\phi_n
=k_\pm\pi$ for any integers $k_+$, $k_-$. Only the vacua and (anti)kink
are stable, however. We shall assume $0<h<2$, so the results apply to the
dipole chain if we choose $h=2/\sqrt{3}$ and apply (\ref{sym}).

It is helpful to think of the static field equation as the variational
equation for the functional $E_P$. To be precise, for any $k_-,k_+\in\Z$, 
let
\beq
\ell^2_{k_-,k_+}:=\{\phi:\Z\ra\R\, |\, \sum_{n=-1}^{-\infty}(\phi_n-k_-\pi)^2
+\sum_{n=0}^\infty(\phi_n-k_+\pi)^2<\infty\}.
\eeq 
Note that $\ell^2_{0,0}=\ell^2$, the Hilbert space of square summable
sequences, and $\ell_{k_-,k_+}$ is an affine space modelled on
$\ell^2$, whose elements have boundary behaviour $\lim_{n\ra\pm\infty}\phi_n
=k_\pm\pi$. It is straightforward to verify
 that $E_P:\ell^2_{k_-,k_+}\ra \R$
is $C^2$, and that $\phi$ is a critical point of $E_P$ if and only if
each of its constituent triples $(\phi_{n-1},\phi_n,\phi_{n+1})$ satisfies
the static field equation
\beq\label{stat}
\frac{\cd E_P}{\cd\phi_n}=
-\frac{4-h^2}{8h}\cos\phi_{n}(\sin\phi_{n+1}+\sin\phi_{n-1})+
\frac{4+h^2}{8h}\sin\phi_{n}(\cos\phi_{n+1}+\cos\phi_{n-1})=0.
\eeq
So critical points of $E_P$ on $\ell^2_{k_-,k_+}$ are static solutions of
the model with $\lim_{n\ra\pm\infty}=k_\pm\pi$, and vice versa. 
Note that {\em any} triple with $\phi_{n+1}-\phi_{n-1}=\pi\, \mod 2\pi$
automatically satisfies (\ref{stat}) irrespective of its central
value $\phi_n$, by $\pi$-antiperiodicity of $\cos$ and $\sin$. 
Such triples turn out to be important for our analysis of static solutions,
so we make the following definition:

\begin{defn} A consecutive triple $(\phi_{n-1},\phi_n,\phi_{n+1})$ in a
sequence $\phi$ is called {\em exceptional} if $\phi_{n+1}-\phi_{n-1}=
(2k+1)\pi$ for some $k\in\Z$.
\end{defn}

To discuss stability of static solutions, we must introduce the second 
variation, or Hessian, of $E_P$. Let $\phi$ be a critical point of $E_P$ and
$\phi^{s,t}$ be a two-parameter variation of it, that is, a smooth map
$(-\epsilon,\epsilon)\times(-\epsilon,\epsilon)\ra\ell^2_{k_-,k_+}$ with
$\phi^{0,0}=\phi$. Let $u=\cd_s\phi^{s,t}|_{(0,0)}, 
v=\cd_t\phi^{s,t}|_{(0,0)}\in\ell^2$. Then the Hessian of $E_P$ at $\phi$
is the symmetric bilinear form $\hess_\phi:\ell^2\times\ell^2\ra\R$
defined by
\beq
\hess_\phi(u,v)=
\left.\frac{\cd^2\:\: }{\cd s\cd t}\right|_{s=t=0}E_P[\phi^{s,t}].
\eeq
We say that $\phi$ is {\em stable} if the associated quadratic form
$u\mapsto\hess_\phi(u,u)$ is positive semi-definite
(i.e.\ $\hess_\phi(u,u)\geq 0$ for all $u$), 
and {\em unstable} otherwise. This definition is motivated by the
analogy with a point particle moving in $\R^3$ under the influence of a
potential $V$, since then saddle points and maxima of $V$ are clearly
unstable equilibria.
We will consider the linear stability criterion later in this section.
By considering the variations $\phi^{s,t}=\phi+(s+t)u$ and
$\phi^t=\phi+tu$, one sees that
\beq
\hess_\phi(u,u)=\left.\frac{d^2 E_P[\phi^t]}{dt^2}\right|_{t=0},
\eeq
so to prove instability it suffices to exhibit a one-parameter variation
$\phi^t$ of $\phi$ such that $d^2E_P[\phi^t]/dt^2<0$ at $t=0$. 

Let us briefly recall the construction of the kink solutions of
the model (\ref{kink}). 
The key idea is to write $E_P$ in the form
\beq
E_P[\phi]=\frac{h}{4}\sum_n(D_n^2+F_n^2),\quad
D_n:=\frac{2}{h}\sin\frac{1}{2}(\phi_{n+1}-\phi_n),\quad
F_n:=\sin\frac{1}{2}(\phi_{n+1}+\phi_n),
\eeq
and observe that $D_nF_n=-\Delta\cos\phi_n$, where $\Delta f_n:=
(f_{n+1}-f_n)/h$. Hence
\beq\label{bb}
0\leq\frac{h}{4}\sum_n(D_n-F_n)^2=E_P+\frac{h}{2}\sum_n\Delta\cos
\phi_n=E_P+\frac{1}{2}(\cos k_+\pi-\cos k_-\pi).
\eeq
It follows that
$E_P\geq 1$ on $\ell^2_{0,1}$, and $E_P=1$ if
and only if $D_n=F_n$ for all $n$. This discrete Bogomol'nyi equation may
be written
\beq\label{dbe}
\tan\frac{\phi_{n+1}}{2}=\frac{2+h}{2-h}\tan\frac{\phi_n}{2},
\eeq
and its general non-vacuum solution is (\ref{kink}). 
On $\ell^2_{1,0}$, a similar argument beginning with summand $(D_n+F_n)^2$
shows that $E_P\geq 1$ and $E_P=1$ if and only if $D_n=-F_n$ for all
$n$, that is,
\beq\label{adbe}
\tan\frac{\phi_{n+1}}{2}=\frac{2-h}{2+h}\tan\frac{\phi_n}{2},
\eeq
whose general non-vacuum solution is
\beq\label{antikink}
\phi_n=2\tan^{-1}e^{-a(n-b)}.
\eeq
We shall refer to (\ref{kink}), (\ref{antikink}) and the vacua as
Bogomol'nyi solutions of the TDSG system. Since they globally minimize
$E_P$ within their boundary class, they are automatically critical points
of $E_P$ with non-negative Hessian, and hence stable static
solutions of the model.

To construct {\em non-Bogomol'nyi} static solutions, we exploit
the possibility that $\phi$ may have exceptional triples. An interesting
fact about such a triple is that the potential energy contributed by its
pair of constituent links is independent of the values of the triple.

\begin{lemma}\label{indep} The potential energy contributed by any
exceptional triple $(\phi_{n-1},\phi_n,\phi_{n+1})$ is $\frac{4+h^2}{4h}$
\end{lemma}

\noindent
{\it Proof:} Since $\phi_{n+1}=\phi_{n-1}+(2k+1)\pi$ for some $k\in\Z$, the
energy of the pair of links $(\phi_{n-1},\phi_n),(\phi_n,\phi_{n+1})$ is
\bea
\frac{h}{4}(D_{n-1}^2+D_n^2+F_{n-1}^2+F_{n}^2)&=&
\frac{1}{h}[\sin^2\frac{1}{2}(\phi_n-\phi_{n-1})
+\sin^2\frac{1}{2}(\phi_{n-1}+\pi-\phi_n)]\nonumber \\
&&+
\frac{h}{4}[\sin^2\frac{1}{2}(\phi_n+\phi_{n-1})
+\sin^2\frac{1}{2}(\phi_{n-1}+\pi+\phi_n)]=\frac{4+h^2}{4h}\nonumber
\eea
\begin{flushright}$\Box$\end{flushright}
\vspace{0.5cm}

\noindent We now have

\begin{prop} \label{shsc1}
Let $\phi\in\ell^2_{k_-,k_+}$ be a static solution with 
energy $E$ and $k$ be an integer. Then $\bar{\phi}\in\ell^2_{k_-,
k_++1}$ defined by
$$
\bar{\phi}_n=\left\{\begin{array}{ll}
{\phi}_n& n\leq k\\
{\phi}_{n-2}+\pi& n\geq k+1\end{array}\right.
$$
is also a static solution, with energy 
$\bar{E}=E+\frac{4+h^2}{4h}$.
\end{prop}

\noindent {\it Proof:}
Since $\phi$ is a static solution, every triple 
$(\bar{\phi}_{n-1},\bar{\phi}_n,\bar{\phi}_{n+1})$  with $n=\notin\{k,k+1\}$
satisfies (\ref{stat}), while  $\bar{\phi}_{k+1}-\bar{\phi}_{k-1}=
\bar{\phi}_{k+2}-\bar{\phi}_k=\pi$ by construction, so the triples
centred on $n=k$ and
$n=k+1$ are exceptional. Hence $\bar{\phi}$ is a static solution.

To compute the energy of $\bar{\phi}$ note that 
$D_n(\bar{\phi})^2+F_n(\bar{\phi})^2=D_n(\phi)^2+F_n(\phi)^2$ for
all $n<k$ and $D_n(\bar{\phi})^2+F_n(\bar{\phi})^2=
D_{n-2}(\phi)^2+F_{n-2}(\phi)^2$ for $n>k+1$. Hence $E_P[\bar{\phi}]=
E_P[\phi]+\delta E$, where $\delta E$ is the energy of the pair of links
forming the exceptional triple 
$(\bar{\phi}_{k-1},\bar{\phi}_k,\bar{\phi}_{k+1})$. But $\delta E=(4+h^2)/4h$
by Lemma \ref{indep}, so
\beq
E_P[\bar{\phi}]=E_P[{\phi}]+\frac{4+h^2}{4h},
\eeq
as was to be proved.\hspace*{\fill} $\Box$
\vspace{0.5cm}

\begin{figure}[htb]
\begin{center}
\begin{tabular}{ccc} 
(a)&
\raisebox{-5ex}{\includegraphics[scale=0.42]{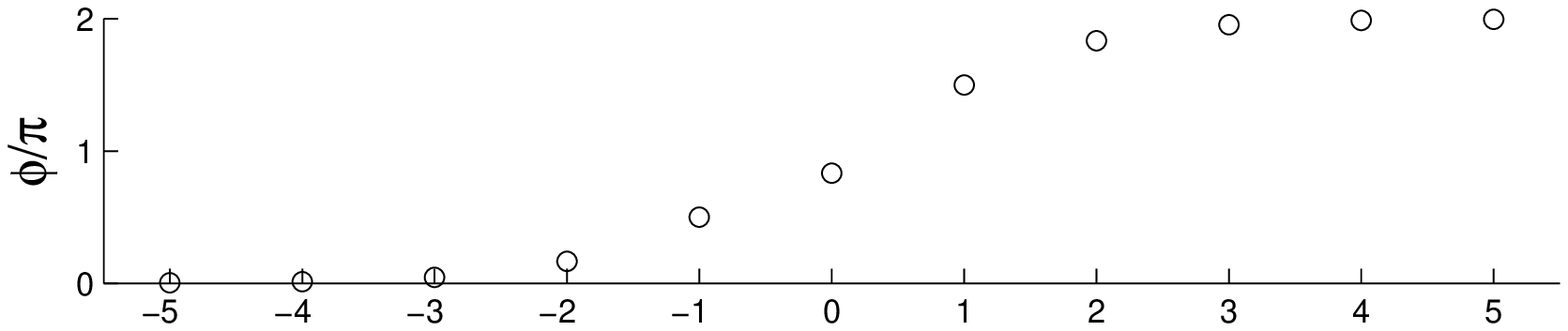}}&
\raisebox{-5ex}{\includegraphics[scale=0.42]{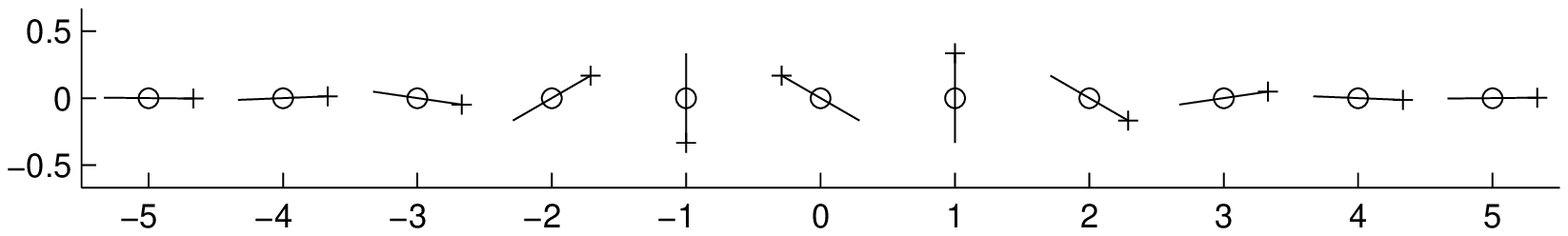}}\\
(b)&
\raisebox{-5ex}{\includegraphics[scale=0.42]{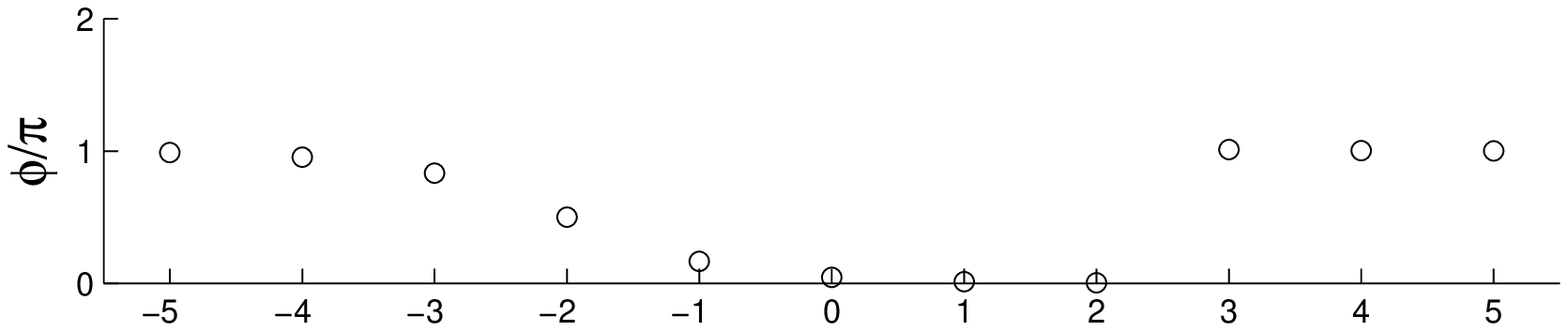}}&
\raisebox{-5ex}{\includegraphics[scale=0.42]{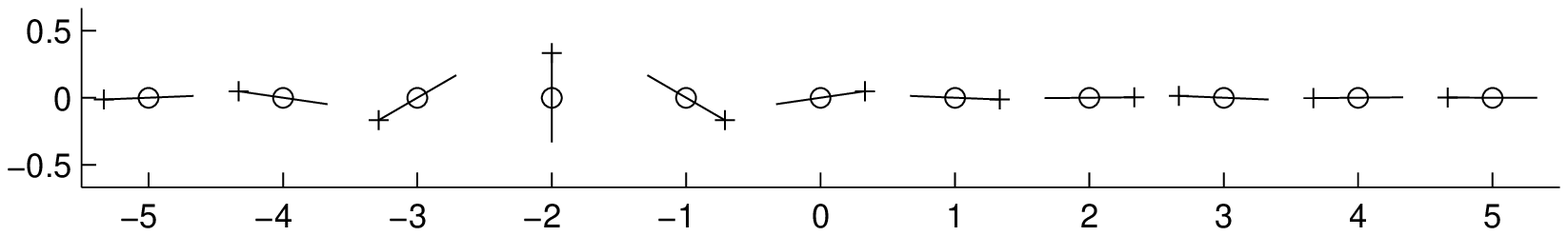}}\\
(c)&
\raisebox{-8ex}{\includegraphics[scale=0.42]{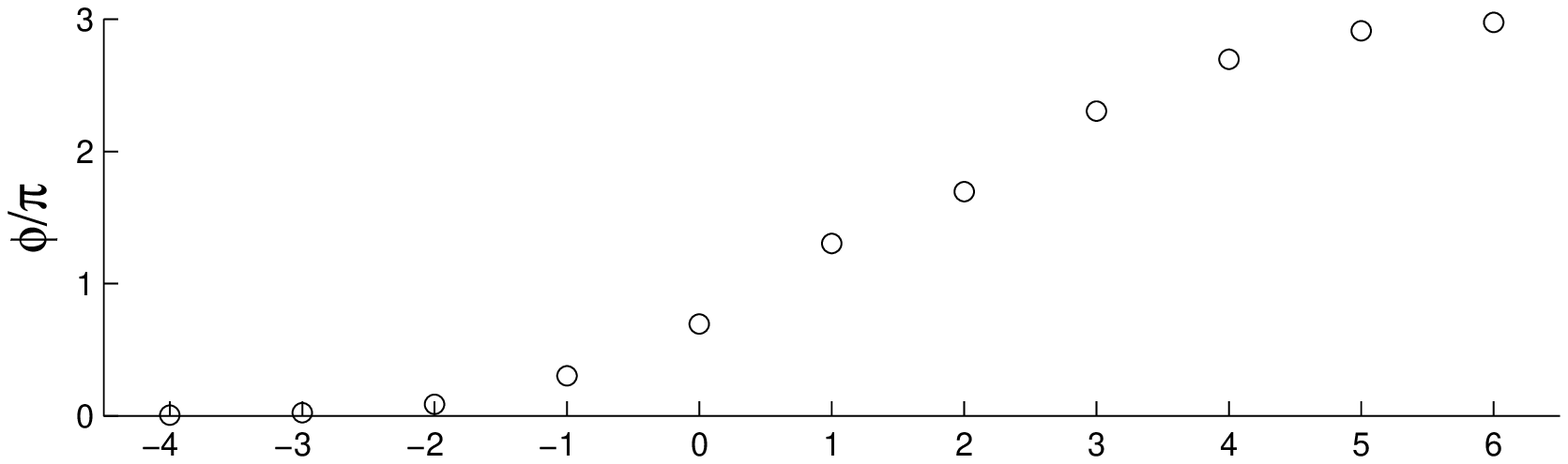}}&
\raisebox{-8ex}{\includegraphics[scale=0.42]{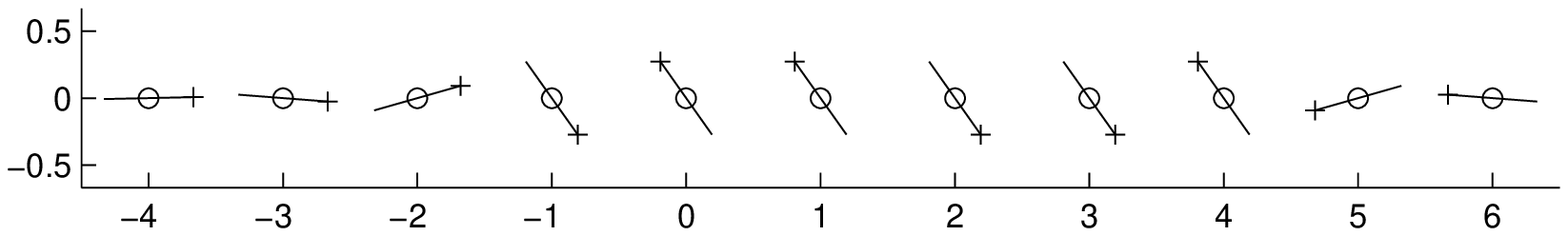}}\\
\end{tabular}
\end{center}
\caption{Some non-Bogomol'nyi static solutions of the spacing 
$h_*=2/\sqrt{3}$ TDSG system (left), and their corresponding dipole 
multikinks (right): a double kink (a), a kink-antikink (b), and a
triple kink (c).}\label{fig2}
\end{figure}

If we choose ${\phi}=0$, the vacuum, in Proposition \ref{shsc1},
 then $\bar{\phi}$ is a step-function kink located
at $k$, of energy $(4+h^2)/4h>1$. 
If ${\phi}$ is a kink, $\bar{\phi}$
is a double kink whose energy exceeds $2$, twice the energy of a single
kink. If ${\phi}$ is an antikink, $\bar{\phi}$ is a (step) kink-antikink
pair. Clearly we can iterate the procedure to generate non-Bogomol'nyi
solutions with arbitrary boundary values. All solutions so
constructed have higher energy
than the minimum within $\ell^2_{k_-,k_+}$ ($1$ if $k_+-k_-$ is odd,
$0$ otherwise), leading one to expect them to be unstable. Note also that
the energy of any such solution diverges as $h\ra 0$, as one would expect
given that the continuum sine-Gordon model has no analogous static solutions.
A static double kink, kink-antikink and triple kink are depicted in figure
\ref{fig2}.

In fact, one can prove that {\em all}\, non-Bogomol'nyi static solutions are
unstable, independent of any particular construction. Crucial to the
argument is the fact that every finite energy static solution has
Bogomol'nyi ``tails'' as $n\ra\pm\infty$. More precisely:

\begin{prop}\label{tail} Let $\phi$ be a finite energy static solution.
If $\phi$ has no exceptional triples, it is either a vacuum, a kink or
an antikink. If $\phi$ has one or more exceptional triples,
there exists $N>0$ such that $D_n=F_n$ for all $n>N$ or 
$D_n=-F_n$ for all $n>N$, and $D_n=F_n$ for all $n<-N$ or $D_n=-F_n$
for all $n<-N$. Hence $\phi$ coincides with a vacuum, a kink
or an antikink for $n>N$, and likewise for $n<-N$.
\end{prop}

\noindent
{\it Proof:} The proof rests on a pair of factorizations due to Ward
\cite{war}. First,
\beq\label{eq0}
\frac{4}{h}\frac{\cd E_P}{\cd \phi_n}=-\cos\frac{1}{2}(\phi_{n+1}-\phi_{n-1})
\left[\frac{4}{h^2}\sin\frac{1}{2}(\phi_{n+1}-2\phi_n+\phi_{n-1})-
\sin\frac{1}{2}(\phi_{n+1}+2\phi_n+\phi_{n-1})\right],
\eeq
so every {\em unexceptional} triple in the static solution
$\phi$ satisfies
\beq\label{eq1}
\frac{4}{h^2}\sin\frac{1}{2}(\phi_{n+1}-2\phi_n+\phi_{n-1})-
\sin\frac{1}{2}(\phi_{n+1}+2\phi_n+\phi_{n-1})=0.
\eeq
Second,
\beq\label{eq1.5}
\Delta(D_{n-1}^2-F_{n-1}^2)=\sin\frac{1}{2}(\phi_{n+1}-\phi_{n-1})
\left[\frac{4}{h^2}\sin\frac{1}{2}(\phi_{n+1}-2\phi_n+\phi_{n-1})-
\sin\frac{1}{2}(\phi_{n+1}+2\phi_n+\phi_{n-1})\right].
\eeq
Now since $E_P[\phi]$ is finite, $\phi$ may have at most finitely
many exceptional triples, by Lemma \ref{indep}, so there exists
$N>0$ such that no triple $(\phi_{n-1},\phi_n,\phi_{n+1})$ is exceptional
for $|n|>N$. But then for all $n>N$, (\ref{eq1}) holds, so (\ref{eq1.5}) 
implies that $D_{n-1}^2-F_{n-1}^2=C$, a constant. But 
$\sum_n(D_n^2+F_n^2)<\infty$, so $D_n,F_n\ra 0$ and we conclude that 
$D_n^2=F_n^2$ for all $n>N$. Similarly $D_n^2=F_n^2$ for all $n<-N$. If
$\phi$ has no exceptional triples, the same argument
shows that $D_n^2=F_n^2$ for all $n$.

It remains to rule out the possibility (for $n>N$) that
$D_{n-1}=F_{n-1}$ while $D_n=-F_n$, or 
$D_{n-1}=-F_{n-1}$ while $D_n=F_n$, for we can then conclude that
$\phi$ uniformly satisfies
either (\ref{dbe}) or (\ref{adbe}) on its right tail. Assume the
contrary. Then 
\beq
\tan\frac{\phi_{n+1}}{2}
=\tan\frac{\phi_{n-1}}{2}
\eeq
so $\phi_{n-1}=\phi_{n+1}\, \mod 2\pi$. But this contradicts
(\ref{stat}), unless $\phi_{n-1}=\phi_{n+1}=0$ or $\pi\, \mod 2\pi$.
In this case, the right tail of $\phi$ is a vacuum since
$0$ and $\pi$ are fixed points of both (\ref{dbe}) and (\ref{adbe}). Hence
 either $\phi$ satisfies (\ref{dbe}) for all $n>N$, or
it satisfies (\ref{adbe}) for all $n>N$, or both (the vacua). The left tail,
$n<-N$, is handled similarly. Again, if $\phi$ has no exceptional
triples then either $D_n=F_n$ for all $n$, or $D_n=-F_n$ for all $n$,
so $\phi$ is a kink, antikink or vacuum.\hfill$\Box$
\vspace{0.5cm}

\begin{thm}\label{unstable}
 Let $\phi$ be a  finite energy static 
solution other than the kink, antikink or vacuum. Then 
$\phi$ is unstable, in the sense that there exists $u\in\ell^2$ such that
$\hess_\phi(u,u)<0$.
\end{thm}

\noindent
{\it Proof:} 
By Proposition \ref{tail}, $\phi$ has at least one and at most finitely
many exceptional triples
We may assume, without loss of generality, that the leftmost
exceptional triple
is $(\phi_{-1},\phi_0,\phi_1)$, so that $D_n^2=F_n^2$ for $n\leq -1$.
Let $\phi^{\alpha,t}$ be the 1-parameter family (parametrized by $\alpha$) 
 of variations (parametrized by $t$) of $\phi$, defined by
\beq
\phi_n^{\alpha,t}=\left\{\begin{array}{ll}
\phi_n& n\notin\{-1,0\}\\
\phi_{-1}+\alpha t& n=-1\\
\phi_0+t & n=0,\end{array}\right.
\eeq
 (note that $\phi^{\alpha,0}\equiv\phi$ for
all $\alpha$), and let $u^\alpha=d\phi^{\alpha,t}/dt|_{t=0}\in\ell^2$. 
We will show that there exists an $\alpha_*$ such that
\beq
\hess_\phi(u^{\alpha_*},u^{\alpha_*})=
\left.\frac{d^2\,\,}{dt^2}\right|_{t=0}E_P[\phi^{\alpha_*,t}]<0.
\eeq
Note that $u^\alpha=\alpha e_{-1}+e_0$, where $\{e_n\, |\, n\in\Z\}$ is the 
standard basis for $\ell^2$, so 
\beq
\hess_\phi(u^\alpha,u^\alpha)=a_0+a_1\alpha+a_2\alpha^2=:p_2(\alpha),
\eeq
some quadratic polynomial in $\alpha$.

Now, $\phi^{0,t}$ only
varies $\phi$ by changing $\phi_0$, which affects only the energies
of the links $(\phi_{-1}^{0,t},\phi_0^{0,t})$ and 
$(\phi_0^{0,t},\phi_{1}^{0,t})$. But these two links
constitute an exceptional triple for all $t$, so the sum of 
their energies is
$(4+h^2)/4h$, independent of $t$ by Lemma \ref{indep}. Hence $E_P[\phi^{0,t}]
=E_P[\phi]$ for all $t$, so
\beq
a_0=\hess_\phi(u^0,u^0)=
\left.\frac{d^2\,\,}{dt^2}\right|_{t=0}E_P[\phi^{0,t}]=0,
\eeq
and $p_2(\alpha)=
a_1\alpha+a_2\alpha^2$. It thus suffices to show that $p_2(-1)\neq p_2(1)$,
since then $a_1\neq 0$ so $p_2(\alpha)$ takes  negative values close to
$0$. 

The variation $\phi^{\alpha,t}$ changes only $\phi_{-1}$ and $
\phi_0$, so
\bea
E_P[\phi^{\alpha,t}]&=&{\rm const}-
\frac{1}{2h}\left\{\cos(\phi_{-1}-\phi_{-2}+\alpha t)+
\cos(\phi_0-\phi_{-1}+(1-\alpha)t)-\cos(\phi_0-\phi_{-1}+t)\right\}\nonumber
\\
&&\quad
-\frac{h}{8}\left\{\cos(\phi_{-1}+\phi_{-2}+\alpha t)+
\cos(\phi_0+\phi_{-1}+(1+\alpha)t)-\cos(\phi_0+\phi_{-1}+t)\right\}
\nonumber \\
\Rightarrow
p(1)&=&
\frac{1}{2h}\left\{\cos(\phi_{-1}-\phi_{-2})
-\cos(\phi_0-\phi_{-1})\right\}
+\frac{h}{8}\left\{\cos(\phi_{-1}+\phi_{-2})+
3\cos(\phi_0+\phi_{-1})\right\},\nonumber \\
p(-1)&=&
\frac{1}{2h}\left\{\cos(\phi_{-1}-\phi_{-2})+
3\cos(\phi_0-\phi_{-1})\right\}
+\frac{h}{8}\left\{\cos(\phi_{-1}+\phi_{-2})
-\cos(\phi_0+\phi_{-1})\right\}\nonumber \\
\Rightarrow 2a_1&=&p(1)-p(-1)=-\frac{2}{h}\cos(\phi_0-\phi_{-1})+
\frac{h}{2}\cos(\phi_0+\phi_{-1}).
\eea
But recall that $D_{-1}^2=F_{-1}^2$, which implies that
\beq
\cos(\phi_{0}-\phi_{-1})=1+\frac{h^2}{4}[\cos(\phi_0+\phi_{-1})-1],
\eeq
and hence
\beq\label{a1}
a_1=-\frac{4-h^2}{4h}< 0.
\eeq
Thus $\hess_\phi(u^\alpha,u^\alpha)<0$ for $\alpha$ sufficiently small and
positive, so $\phi$ is unstable.\hfill$\Box$ 
\vspace{0.5cm}

We will now deduce from Theorem \ref{unstable} that
all non-Bogomol'nyi static solutions are {\em linearly} unstable. 
If the configuration space of our model were finite dimensional, this would
follow immediately. However, in the infinite dimensional case it is
necessary to proceed carefully.
Let $J^\phi:\ell^2\ra\ell^2$ be the Jacobi operator associated with the
bilinear form $\hess_\phi$, that
is, the unique linear map such that
\beq
\langle u,J^\phi v\rangle_{\ell^2}\equiv\hess_\phi(u,v).
\eeq
Then the linearization of (\ref{eom}) about the static solution $\phi$ may
be thought of as a second order linear ODE on $\ell^2$, namely,
\beq\label{linear}
\frac{h}{2}\ddot{v}+J^\phi v =0.
\eeq
We think of $v(t)\in\ell^2$ as a perturbation of $\phi$. If there
exist perturbations which grow without bound (in $\ell^2$ norm), we
say that $\phi$ is linearly unstable. Such a perturbation certainly
exists if $J^\phi$ has a negative {\em eigenvalue}, 
since we can perturb in the
direction of the corresponding eigenvector. It is not hard to
show from Theorem \ref{unstable}
that $\spec J^\phi$ contains at least one negative real number. However,
we cannot conclude immediately that this number is an eigenvalue, since 
in infinite dimensions it may lie in the continuous or residual part of
$\spec J^\phi$. A more detailed analysis of $\spec J^\phi$ is required.
We begin with two lemmas giving basic properties of $J^\phi$.

\begin{lemma}\label{jsa} $J^\phi$ is bounded and self-adjoint.
\end{lemma}

\noindent
{\it Proof:} Clearly $J^\phi$ is symmetric by definition, so 
self-adjointness will follow from boundedness. As before, 
let $\{e_n\, |\, n\in\Z\}$
be the usual basis for $\ell^2$ and $J^\phi_{nm}:=\langle e_n,J^\phi e_m
\rangle$. Then 
\bea
J^\phi_{nm}&=&\left.\frac{\cd^2E_P}{\cd\phi_n\cd\phi_m}\right|_\phi\nonumber
\\
&=& \frac{h}{8}\bigg\{\delta_{n,m}\left[\frac{4}{h^2}(\cos(\phi_m-\phi_{m-1})
+\cos(\phi_m-\phi_{m+1}))+\cos(\phi_m+\phi_{m-1})+\cos(\phi_m+\phi_{m+1})
\right]\nonumber \\
&&+(\delta_{n,m-1}+\delta_{n,m+1})\left[-\frac{4}{h^2}\cos(\phi_m-\phi_n)+
\cos(\phi_m+\phi_n)\right]\bigg\}.
\label{d2E}
\eea
Hence $(J^\phi v)_n=c_nv_n+d_{n-1}v_{n-1}+d_{n+1}v_{n+1}$, 
where $c_n$, $d_n$ are
sequences bounded independent of $\phi$: $|c_n|,|d_n|\leq (4+h^2)/4h$. Thus
\bea
||J^\phi v||^2&=&\sum_n(c_nv_n+d_{n-1}v_{n-1}+d_{n+1}v_{n+1})^2\nonumber \\
&\leq& 3\left(\frac{4+h^2}{4h}\right)^2\sum_n(v_n^2+v_{n-1}^2+v_{n+1}^2)
=9\left(\frac{4+h^2}{4h}\right)^2||v||^2,\nonumber
\eea
as was to be proved.\hfill $\Box$
\vspace{0.5cm}

It follows immediately from Lemma \ref{jsa} that $\spec J^\phi$ 
is real, bounded
and consists of eigenvalues and continuous spectrum only 
(there is no residual
spectrum) \cite{chodewmor}.

\begin{lemma}[Asymptotic positivity]\label{local}
Let $\phi$ be a finite energy static solution, and for each $N\in\Z^+$,
let $V_N=\{v\in\ell^2\, |\, \mbox{$v_n=0$ for all $|n|> N$}\}$. Then
for all $\eps>0$, there exists $N$ such that
$$
\langle v,J^\phi v\rangle \geq (\frac{h}{2}-\eps)||v||^2
$$
for all $v\in V_N^\perp$. 
\end{lemma}

\noindent
{\it Proof:} By Proposition \ref{tail}, there exists $N_1>0$ such that
$\phi$ coincides with a vacuum or (anti)kink located at $b_+$ for $n>N_1$,
and a vacuum or (anti)kink located at $b_-$ for $n<-N_1$. 
Given the exponential decay properties of (anti)kinks 
(\ref{kink}),(\ref{antikink}),
it follows that
\beq\label{shgh2}
|\cos(\phi_{n+1}\pm\phi_n)-1|<Ce^{-2a|n|}
\eeq
for all $|n|>N_1+1$, where $C>0$ depends on $b_\pm$ and $h$, but not $N_1$.
(Recall that $a=\log((2+h)/(2-h))$.) 
Let $J^0$ be the Jacobi operator of the vacuum,
\beq
J^0v=\frac{4+h^2}{4h}v-\frac{4-h^2}{8h}(\tau_+v+\tau_-v),
\eeq
where $\tau_\pm$ are the unit shift maps, $(\tau_\pm v)_n=v_{n\pm 1}$, and
$L=J^\phi-J^0$. 

Choose $\eps>0$, and let $N$ be any integer exceeding both
$N_1$ and $\frac{1}{2a}
\log\left(\frac{2C}{h\eps}\right)$.
Then for all $v\perp V_N$, $v_n=0$ for $|n|\leq N$, so
we see from (\ref{d2E}) and (\ref{shgh2}) that
\bea
|\langle v,L v\rangle|&\leq&\sum_{m<-N}\left\{
\frac{4+h^2}{4h}Ce^{-2aN}v_m^2+\frac{4-h^2}{4h}|v_mv_{m-1}|\right\}\nonumber
\\
&&+
\sum_{m>N}\left\{
\frac{4+h^2}{4h}Ce^{-2aN}v_m^2+\frac{4-h^2}{4h}|v_mv_{m+1}|\right\}\nonumber
\\
&=&\left\{\frac{4+h^2}{4h}||v||^2+\frac{4-h^2}{4h}\sum_{m\in\Z}|v_mv_{m+1}|
\right\}Ce^{-2aN}\nonumber \\
&\leq&\frac{2}{h}Ce^{-2aN}||v||^2<\eps||v||^2\nonumber
\eea
since $2|v_mv_{m+1}|\leq v_m^2+v_{m+1}^2$. It follows, therefore, that
\bea
\langle v,J^\phi v\rangle &=& \langle v,J^0v\rangle+\langle v,Lv\rangle
\geq \langle v,J^0v\rangle - |\langle v,Lv\rangle| \nonumber \\
&\geq& \langle v,J^0v\rangle - \eps ||v||^2 
=\frac{4+h^2}{4h}||v||^2-\frac{4-h^2}{2h}\langle v,\tau_+v\rangle
- \eps ||v||^2 \nonumber \\
&\geq&\frac{h}{2}||v||^2-\eps||v||^2\nonumber
\eea
since $\tau_-^*=\tau_+$, $\langle v,\tau_+\rangle\leq ||v||\, 
||\tau_+v||$ by Cauchy-Schwarz, and $\tau_+$ is an isometry.
\hfill $\Box$
\vspace{0.5cm}

The key point is that $J^\phi$ is positive on a subspace of finite
codimension (namely $V_N^\perp$), from which we will see that it can
have no negative continuous spectrum. Therefore, if $J^\phi$
 has some negative spectrum, as it
does when $\phi$ is non-Bogomol'nyi, it must have at least one negative
eigenvalue.
We are now ready to state and prove the required spectral result.

\begin{thm}\label{spec}
Let $\phi$ be a  finite energy static 
solution other than the kink, antikink or vacuum. Then $J^\phi$ has
a negative eigenvalue of finite multiplicity.
\end{thm}

\noindent
{\it Proof:} Since $J$ (we will drop the superscript)
is bounded and self adjoint, its spectrum is
contained in the real interval $[Q_1,Q_2]$ where
$$
Q_1:=\inf_{v\neq 0}\frac{\langle v,J v\rangle}{||v||^2},\qquad
Q_2:=\sup_{v\neq 0}\frac{\langle v,J v\rangle}{||v||^2},
$$
and, furthermore, contains $Q_1$ and $Q_2$ \cite{tay}. By Theorem
\ref{unstable}, $Q_1<0$, so
it remains to show that $Q_1$ is an eigenvalue  of finite multiplicity. 

Choose
$\eps\in (0,h/2)$. Then by
Lemma \ref{local}, there exists $N>0$ such that $\langle v,J v\rangle
\geq (\frac{h}{2}-\eps)||v||^2$ 
for all $v\in V_N^\perp$. Let $P:\ell^2\ra\ell^2$ be 
orthogonal projection onto $V_N$, $P':=1-P$, $T:=P'JP'$ and
$A=P'JP+PJP'+PJP$. Then $J=T+A$, both $T$ and $A$ are bounded and
self-adjoint, and $T$ is, by construction, positive. Given any 
bounded sequence
$u^i$ in $\ell^2$, $Au^i$ is bounded, and takes values in a finite
dimensional subspace of $\ell^2$ (namely $V_N\oplus P'J(V_N)$), so has
a convergent subsequence by the Bolzano-Weierstrass theorem. Thus
$A$ is compact, and hence trivially $T$-relatively compact
\cite[p194]{kat}. Thus ${\rm spec}_e\, J$, the {\em essential} spectrum of 
$J=T+A$ 
coincides with the essential spectrum of $T$ \cite[p244]{kat},
which is bounded below by $\frac{h}{2}-\eps$ \cite{tay}. 
Hence $Q_1$ is not in ${\rm spec}_eJ$. 
But $J$ is
self-adjoint, so $\spec J\backslash {\rm spec}_e\, J$ contains only
eigenvalues of finite multiplicity \cite[p518]{kat}. Hence $Q_1$ is a
negative eigenvalue of finite multiplicity, as was to be proved.
\hfill $\Box$
\vspace{0.5cm}

We note in passing that, since $\eps$ could be chosen freely in the above 
argument, we
get the extra information that the continuous spectrum of $J$ is bounded 
below by $\frac{2}{h}$.

\begin{cor}[Linear instability]
Let $\phi$ be a finite energy static solution other than the kink, antikink
or vacuum. Then the linearized TDSG flow about $\phi$ supports solutions
which grow unbounded exponentially fast.
\end{cor}

\noindent
{\it Proof:}
Let $u\in\ell^2$ be the eigenvector associated with the negative
eigenvalue $Q_1=:-\nu^2$. Then (\ref{linear}) supports the  solution
\beq
v(t)=u\exp(\sqrt{\frac{2}{h}}\nu t),
\eeq
which grows unbounded exponentially fast. \hfill$\Box$
\vspace{0.5cm}

We close this section by noting that there is a
two parameter family of static solutions for which
{\em every} triple is exceptional:
\beq
\phi_n=\left\{\begin{array}{ll}
\alpha& n=0\, \mod 4\\
\beta& n=1\, \mod 4\\
\alpha+\pi& n=2\, \mod 4\\
\beta+\pi& n=3\, \mod 4.\end{array}\right.
\eeq
These stand at the opposite extreme from the Bogomol'nyi solutions.
Clearly they have infinite energy.

\section{Exact propagating kinks}
\label{sec:prop}\news

An important question is whether there exist moving kink solutions
that propagate with constant velocity. In other words, in this
section we are going to look for solutions that satisfy the
following condition:
\beq
\phi_n(t)=\phi(n-st) \equiv \phi(z) \;,\quad z=n-st \;.
\label{eq4}
\eeq
Substituting this ansatz into (\ref{eom}), we see that the profile
$\phi$
satisfies a nonlinear advance-delay ODE, namely
\begin{eqnarray}
\nonumber
s^2\phi''(z)&=&\frac{4-h^2}{4h^2}\cos {\phi(z)} 
\left [\sin {\phi(z+1)}+\sin {\phi(z-1)} \right ]\\
&&-\frac{4+h^2}{4h^2} \sin {\phi (z)} \left [\cos {\phi(z+1)}+
\cos {\phi(z-1)} \right ] \; .
\label{eom2}
\end{eqnarray}
It is very hard to prove rigorous results on the solutions of such ODEs,
and comparatively little is known in general.
Friesecke and Wattis \cite{friwat} have proved the existence of 
propagating pulses in FPU chains with superquadratic intersite potential.
Somewhat closer to the current situation, Iooss and Kirchg\"assner
\cite{iookir} have proved the existence, in linearly 
coupled oscillator chains, of small amplitude travelling
pulses with small oscillatory tails. Neither method developed in these
papers applies directly to the travelling kink problem we seek to address.

We will approach the problem numerically by seeking kink solutions of
(\ref{eom2}) using the {\it pseudo-spectral method}, 
a highly effective tool for finding travelling-wave solutions in lattices
which
has been developed in a number of papers \cite{ef90pla,defw93pd,sze00pd}.
The idea is to write $\phi$ as a basic kink (for example
$2\tan^{-1}e^{az}$) plus a small correction $\delta\phi(z)$, write
down a truncated Fourier series for $\delta\phi(z)$ and hence reduce
(\ref{eom2}) to a large algebraic system for the unknown Fourier coefficients,
which can be solved using the Newton method. 
For technical details see \cite{defw93pd,sze00pd}.
We stress that this method is based not on lattice simulations but on 
iterative techniques and it enables us to find travelling-wave
solutions with any given precision. 

\begin{figure}[htb]
\begin{center}
\includegraphics[scale=0.5,angle=-90]{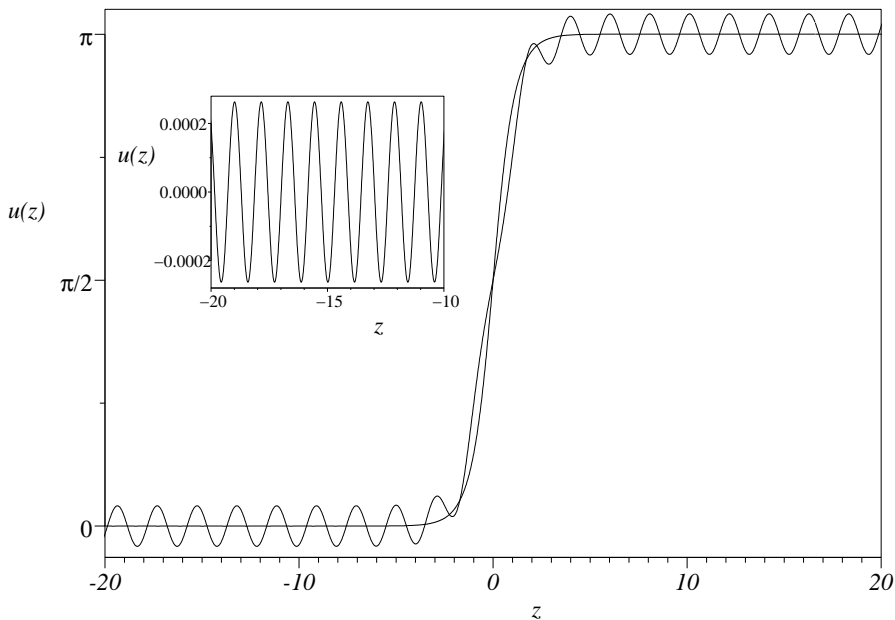}
\end{center}
\caption{Profiles of the nanopteron solution $u(z)$ of (\ref{eom2}) for
$s=0.2$ and $s=0.5$ (with larger oscillating tail). The inset 
shows details of the tail of the
solution for $s=0.2$. In both cases $h=1.3$.}
\label{figy3}
\end{figure}

Straightforward application of the pseudo-spectral method to the TDSG
system shows that, as for conventional
discretizations, moving kink solutions  with oscillating 
tails, or so-called {\it nanopterons}, appear. Two such nanopterons are
depicted in figure \ref{figy3}.
The amplitude $A$ of the oscillating tail depends on speed $s$ and
lattice spacing $h$. If $A(s,h)=0$, one has a genuine travelling kink, since
$\phi$ satisfies kink boundary conditions. Such solutions usually 
have codimension 1 in the $s,h$ parameter space, so they occur at
isolated values of $s$ if $h$ is fixed, and {\it vice versa}.\, A
heuristic explanation of this was provided by Aigner {\it et al} 
\cite{acr03pd}. They think of the travelling kink
profile $\phi$ as a heteroclinic orbit
from $0$ to $\pi$ and informally compute the dimensions of the stable
and unstable manifolds of these fixed points by considering the linearization
of (\ref{eom2}) about them. Every independent oscillatory solution of the
linearization cuts the dimension of each of these manifolds by 1, and 
increases the codimension of the space of travelling kinks by 1.\, 
Applying their argument to the TDSG system leads one to expect that the
codimension equals the number of non-negative solutions $k$ of the
equation
\beq
s^2k^2=1+\frac{4-h^2}{h^2}\sin^2\frac{k}{2},
\eeq
obtained by demanding that $\phi=\cos kz$ be a solution of the linearization
of (\ref{eom2}). 
For fixed $h$, this number is $1$ for $s$ large, but grows without bound
as $s\ra 0$, suggesting that travelling kinks, if they exist at all, should
disappear in the low speed limit. On the other hand, the $s\ra 0$ limit
of the TDSG system is clearly very {\em un}-generic, since the system
supports a continuous orbit of static kinks, so it remains possible
that the arguments of \cite{acr03pd} are misleading in this case.

\begin{figure}[htb]
\begin{center}
\includegraphics[scale=0.5,angle=-90]{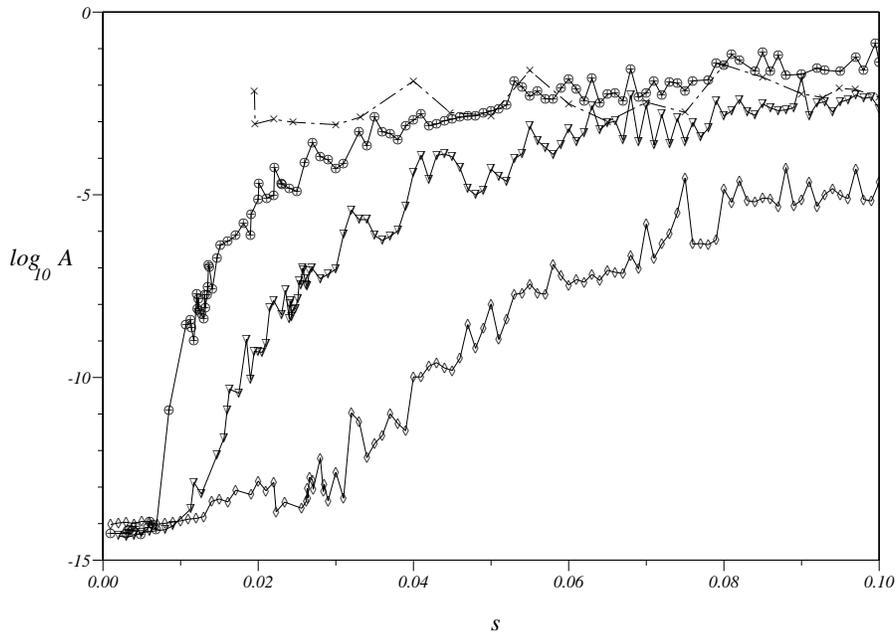}
\end{center}
\caption{Amplitude of the oscillating tail as a function of the 
kink velocity $s$ for  $h=1$ ($\diamond$), $h=1.3$ ($\nabla$), 
$h=1.5$ ($\oplus$)  and
DSG ($\times$) [given by Eq. (\ref{DSG}) with $h=1$]. Connecting
lines are used as guides to the eye.}
\label{figy4}
\end{figure}

To resolve this issue, we have systematically
computed the nanopteron tail amplitude $A$ 
as a function of $s$ for various values of $h$, see figure \ref{figy4}. 
In no case do we see
zeros of $A$, isolated or otherwise,
 at large $s$, so we actually find {\em no} travelling
kinks in the portion of parameter space where they are expected to arise 
with codimension 1.\, The small $s$ regime is more interesting. As $s\ra 0$,
the amplitude drops monotonically to a value so small as to be $0$ to
within numerical tolerance. Certainly $A(0)=0$ since there is no PN barrier,
and $\phi(z)=2\tan^{-1}e^{az}$ solves (\ref{eom2}) at $s=0$.
This small $s$ behaviour is very different from that
 of the conventional discrete sine-Gordon system (DSG), 
\beq
{\ddot \phi}_n=\frac{\phi_{n+1}-2\phi_n+\phi_{n-1}}{h^2}-\sin \phi_n,
\label{DSG}
\eeq 
included in
figure \ref{figy3} for comparison: $A(s)$ clearly
remains bounded away from $0$ for this system.
For the TDSG system, it is possible that $A(s)$ really does
attain the value $0$ on some narrow interval $[0,s_*(h)]$, so that
exact travelling kinks exist for all speeds not exceeding $s_*(h)$.
More likely $A(s)$ rises immediately from $0$. This would be consistent
with some formal asymptotics of Oxtoby {\it et al}
computed in the context of exceptional discretizations of the $\phi^4$
system \cite{oxtpelbar}.
However, the question is numerically inaccessible. All we can say is
that the tail amplitude goes rapidly to zero as $s\ra 0$, and that this
is reflected in a high degree of practical kink mobility, as 
demonstrated in previous studies \cite{spewar,spe2,spe3}, and by our
own simulations, described below.

\begin{figure}[htb]
\begin{center}
\includegraphics[scale=0.5,angle=-90]{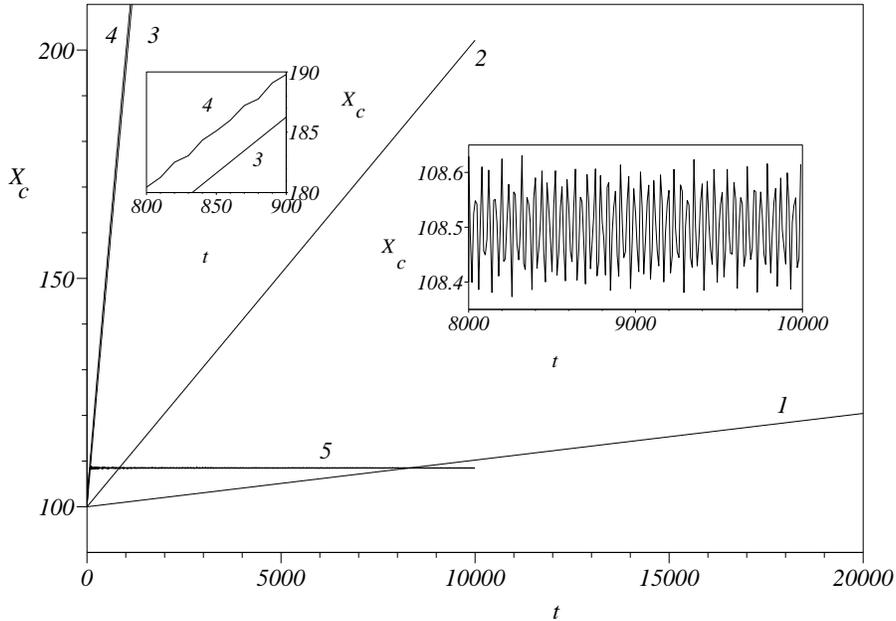}
\end{center}
\caption{Temporal evolution of the kink's centre of mass for
$h=1.3$ for two discretizations of the sine-Gordon equation.
Lines 1, 2, 3 and 4 correspond to the TDSG with 
initial velocities $s=0.001$, $s=0.01$, $s=0.1$ and $s=0.5$, respectively.
Line 5 corresponds to the kink of the conventional DSG
 kicked with initial velocity $s=0.1$.  In the simulations, 
dissipation 
has been introduced at both
ends of the chain. The left inset shows details of lines 3 and 4,
and the right inset shows  details of line 5.}
\label{figy5}
\end{figure}

In order to check the dynamics of the TDSG and DSG systems, we performed
numerical simulations of (\ref{eom}) and (\ref{DSG}). 
The initial conditions were chosen by assuming translational
invariance of the solution. In the case of the TDSG, we substitute
$b=st$ in (\ref{kink}) and compute $\phi_n(0),\dot{\phi}_n(0)$:
\begin{eqnarray}
\phi_n(0)&=&2\tan^{-1}e^{a(n-n_0)},\\
\dot{\phi}_n(0)&=&\frac{d}{dt}\phi_n(t)|_{t=0}=-as/\cosh [a(n-n_0)] \;.
\end{eqnarray}
In the case of the conventional DSG, due to
the absence of an explicit stationary solution, we take its
continuum approximation  $\phi_n(t)=4\tan^{-1}e^{h(n-n_0-st)}$
and compute $\phi_n(0),\dot{\phi}_n(0)$.
Both equations have been simulated using the 4th order Runge-Kutta
method on a finite lattice of size $N$. 
The position of the kink's centre of mass, defined as
\beq
X_c(t)=\frac{\sum_{n=1}^N 
n(\phi_{n+1}-\phi_{n-1})}{2(\phi_N-\phi_1)} \;,
\eeq
has been plotted as a function of time in figure \ref{figy5}. 

One can clearly see from this figure the difference between the two
discretizations. In the TDSG case the kink moves freely along the
lattice even when
kicked with very small initial velocities (lines 1, 2, and 3). 
By inspecting this figure it is easy to see that, for small initial
velocities,  the
actual velocity of propagation almost coincides with the initial 
velocity, so that the kink suffers practically no radiative deceleration.
The figure also shows the existence of a velocity threshold above which
radiative deceleration becomes important: the kink given initial velocity
$s=0.5$ quickly slows down to approximately $s=0.1$ (line 4).
By contrast, line 5 corresponds to the kink of the conventional DSG, kicked 
with 
an initial velocity $s=0.1$.
 The kink propagates several sites and then finds
itself trapped by the lattice. The same happens for smaller
values of the initial velocity $s$.
Thus, in the dynamical picture, the absence of the PN barrier
for the TDSG manifests itself in the possibility, for all
practical purposes, of free
propagation of kinks with an arbitrarily small velocity.

\section{Conclusion}
\label{sec:conc}\news

We have shown that there is a concrete physical system modelled by an
exceptional discretization of the sine-Gordon model, namely a uniform
chain of electric dipoles, modelled by the TDSG system. Previous work
\cite{spewar} shows that such a chain admits a continuous translation
orbit of static kinks, with no PN barrier, and that these kinks are
highly mobile and exhibit strongly continuum-like behaviour (no 
trapping, low radiative deceleration etc.), despite being inherently
very highly discrete. We have constructed static multikink
solutions of the system and proved that all 
finite energy non-Bogomol'nyi static solutions (i.e.\ all except the 
kink, antikink and vacua) are unstable. 

A key step in the analysis of static solutions
was the observation that, whilever the solution
has no exceptional triple, it ``conserves'' the two-point function
$D_n^2-F_n^2$ as $n$ ``evolves'' along the lattice. This is strongly
reminiscent of Kevrekidis's method \cite{kev}, so one might ask 
whether the
TDSG system conserves both energy and momentum (it falls outside the class
of models considered in \cite{dmikevyos}). The answer appears, unfortunately,
to be no. From (\ref{eom}), (\ref{eq0}) and (\ref{eq1.5}) it follows that
\beq
\sum_n\ddot{\phi}_n\tan\frac{1}{2}(\phi_{n+1}-\phi_{n-1})=0
\eeq
for any finite energy solution, 
but there is no obvious way to write this
as a total time derivative.

We have investigated the possibility that the TDSG system admits
exact constant-speed propagating kinks, finding
only {\em nanopterons}, that is, kinks with spatially
oscillatory tails. This is in contrast to the results of
Oxtoby {\it et al} on exceptional discrete $\phi^4$ systems, who find 
isolated velocities for which the tail amplitude vanishes exactly so that
exact propagating kinks exist \cite{oxtpelbar}. 
In the TDSG case, the tail amplitude
vanishes rapidly in the zero velocity limit, and may, in fact, remain exactly
zero on a small velocity interval, but the resolution of
this issue is beyond the
capability of our numerics. If so, the travelling kinks would form a
codimension $0$ family in a regime where, if they exist at all, they
should be isolated, in contradiction to the heuristic
barrier of Aigner {\it et al} \cite{acr03pd}. This is perhaps unlikely
\cite{oxtpelbar},
but should not be discounted altogether given the other highly ungeneric
features of the system. Whether or not exact propagating
solutions exist,
numerical simulations confirm that the
kink can move freely with 
arbitrarily small velocity, so that 
kinks enjoy a high degree of practical mobility, 
in marked contrast with the conventional DSG.\,
It is possible that exact travelling kinks with a structure more
complicated than that accommodated by the ansatz (\ref{eq4}) exist.
This problem requires a separate investigation.

To our knowledge, this is the first known example of an exceptional
discrete system which models a genuine physical system. One could object
that the model takes into account only the interactions between
nearest neighbour dipoles. This is true, although this is a standard 
approximation made when modelling lattice systems. Including longer range
inter-dipole forces will presumably destroy the exact continuous translation
orbit of static kinks and introduce a small PN barrier. Since longer range
forces are much weaker (note the $|\rv|^{-3}$ in (\ref{dipdip})) one
would hope that the PN barrier is so small as to be negligible in practice.
To estimate how small, we have added to $E_P$ the energy due to
next-to-nearest neighbour dipole pairs,
\beq
E_P'=\frac{1}{8}\, \frac{h}{4}\sum_n\left\{\frac{4}{h^2}\sin^2\frac{1}{2}(
\phi_{n+1}-\phi_{n-1})+\sin^2\frac{1}{2}(\phi_{n+1}+\phi_{n-1})\right\},
\quad h:=\sqrt{12},
\eeq
in natural units, and numerically minimized $E_P+E_P'$ for site centred 
and link centred alternating kinks using a gradient flow method. We find that
site centred kinks now have very slightly lower energy. The PN barrier
\beq
\frac{E(b=\frac{1}{2})-E(b=0)}{E(b=0)}<0.009\%.
\eeq
This is absolutely tiny in comparison with conventional discrete systems,
and strongly suggests that the inclusion of only nearest neighbour
interactions in the model is a sensible approximation.

\section*{Acknowledgements}

YZ acknowledges financial support from INTAS through its Young
Scientists Grant, contract number 03-55-1799. JMS would like to thank
Jonathan Partington for helpful conversations.

\end{document}